\newif\ifconf
\conftrue

\newif\ifcomm
\commtrue 

\newif\ifblind
\blindfalse

\newif\ifacm
\acmfalse

\newif\ifusenix
\usenixfalse

\newif\ifacmart
\ifacm   
   \acmarttrue
\else
   \acmartfalse
\fi

\newif\ifs
\strue

\newif\ifhyperref
\hyperreftrue

\ifconf
    \newcommand{\Conf}[1]{#1}
    \newcommand{\TR}[1]{}
    \newcommand{\Journal}[1]{}  
    \newcommand{\OnlyTR}[1]{}   
\else
    \newcommand{\Conf}[1]{}
    \newcommand{\TR}[1]{#1}
    \newcommand{\Journal}[1]{}  
    \newcommand{\OnlyTR}[1]{#1}   
\fi

\ifacm
	\ifacmart
    	\documentclass[10pt,sigconf,letterpaper,anonymous,nonacm]{acmart}
         \AtBeginDocument{%
            \providecommand\BibTeX{{%
            \normalfont B\kern-0.5em{\scshape i\kern-0.25em b}\kern-0.8em\TeX}}}
        \acmSubmissionID{123-A56-BU3}
    \else
      \Conf{
 	     \documentclass[letterpaper,conference]{sig-alternate-10pt}
      }
      \TR{
     	 \documentclass[letterpaper,conference]{sig-alternate}
      }
      \usepackage[
       pass,
      ]{geometry}
	\fi
\else
	\ifusenix  
		\documentclass[letterpaper,twocolumn,10pt]{article}
		\usepackage{usenix2019,epsfig,endnotes}
	\else  
      \Conf{
        \documentclass[conference]{IEEEtran}
        \IEEEoverridecommandlockouts
      }
      \TR{
          \documentclass[journal]{IEEEtran}
      }
  \fi
\fi

\usepackage{graphicx}

    \usepackage[justification=justified]{caption}
    \usepackage[labelformat=simple]{subcaption}  
    \DeclareCaptionLabelSeparator{periodspace}{.\quad}
    \ifconf
    \else
    \fi
\usepackage[noadjust]{cite}

\usepackage{url}

\usepackage{xcolor}
\usepackage{amsmath}
\usepackage{mathtools}
\usepackage{scalerel}
\usepackage{algorithm}
\usepackage[noend]{algpseudocode}
\usepackage{times}
\usepackage{xspace}
\usepackage{balance}
\usepackage{booktabs}  
\usepackage{tabularx}  

\usepackage{stfloats} 

\usepackage{array}
	\newcolumntype{C}[1]{>{\centering\let\newline\\\arraybackslash\hspace{0pt}}m{#1}}

\ifhyperref
    \usepackage[hidelinks]{hyperref} 
    \hypersetup{pdfstartview=FitH,pdfpagelayout=SinglePage}
\fi

\ifacm 

\else 
	\usepackage{amsmath}
	\usepackage{amsthm}  
    
    \newtheoremstyle{boldthm}{}{}{\itshape}{}{\bfseries}{.}{ }{\thmname{#1}\thmnumber{ #2}\thmnote{ (#3)}} 
	\theoremstyle{boldthm}
\fi

\algdef{SE}[DOWHILE]{Do}{doWhile}{\algorithmicdo}[1]{\algorithmicwhile\ #1}%
\algdef{SE}[SUBALG]{Indent}{EndIndent}{}{\algorithmicend\ }%
\algtext*{Indent}
\algtext*{EndIndent}

\ifacmart
\else
  \newtheorem{theorem}{Theorem}
  \newtheorem{corollary}{Corollary}
  
\fi

\ifcomm
	\newcommand{\mycomm}[3]{{\footnotesize{{\color{#2} \textbf{[#1: #3]}}}}}
     \newcommand{\Fmycomm}[3]{\footnote{{{\color{#2} \textbf{[#1: #3]}}} }}
\else
    \newcommand{\mycomm}[3]{}
    \newcommand{\Fmycomm}[3]{}
\fi

\newcommand{\new}[1]{#1}

\ifs
		
	\newcommand{\ReduceVSpace}{} 
    \newcommand{\ReduceVSpaceA}{\vspace{-4pt}} 
	\newcommand{\BigReduceVSpace}{\vspace{-5pt}} 
    \usepackage[inline]{enumitem}
        \setlist{leftmargin=*} 
    \newcommand{\T}[1]{\par\vspace{2pt plus 1pt minus 1pt}\noindent\textbf{#1}} 
    \newcommand{\U}[1]{\par\noindent\textit{#1}} 
		
\else

	\newcommand{\ReduceVSpace}{}
	\newcommand{\BigReduceVSpace}{}
    \usepackage[inline]{enumitem}
    \newcommand{\T}[1]{\par\smallskip\noindent\textbf{#1}} 
    \newcommand{\U}[1]{\par\smallskip\noindent\textit{#1}} 
\fi

\usepackage[inline]{enumitem}
\setlist{leftmargin=*}

\usepackage{comment}

\newcommand{\se}[1]{Sec.~\ref{#1}} 
\newcommand{\fig}[1]{Fig.~\ref{#1}} 

\ifacm 
  \newcommand{\bp}{\begin{proof}}
  \newcommand{\bpo}{ \begin{proof}[Proof Outline] }
  \newcommand{\ep}{\end{proof}}       
      \newcommand{\proofof}[1]{\begin{proof}[Proof of #1]} 
\else 
    \ifusenix 
      \newcommand{\bp}{\begin{proof}}
      \newcommand{\bpo}{ \begin{proof}[Proof Outline] }
      \newcommand{\ep}{\end{proof}}  
          \newcommand{\proofof}[1]{\begin{proof}[Proof of #1]} 
    \else 
      \newcommand{\bp}{\begin{IEEEproof}}     
      \newcommand{\bpo}{ \begin{IEEEproof}[Proof Outline] }
      \newcommand{\ep}{\end{IEEEproof}}       
          \newcommand{\proofof}[1]{\begin{IEEEproof}[Proof of #1]} 
    \fi
\fi

\newcommand{\ignore}[1]{}

\newcommand{\be}{\begin{equation}}
\newcommand{\ee}{\end{equation}}

\newcommand\MyIncludeGraphics[2][]{
    \IfFileExists{#2}{%
        \includegraphics[#1]{#2}%
    }{%
        \missingfigure[figwidth=7.0cm]{Missing #2}%
    }%
}%

\providecommand{\vs}{{vs.}\xspace}
\providecommand{\ie}{{i.e.,}\xspace}
\providecommand{\eg}{{e.g.,}\xspace}

\newcommand{\newVar}[2]{\newcommand{#1}{\ensuremath{#2}\xspace}}
  \newVar{\server}{S}
  \newVar{\client}{C}
  \newVar{\rclient}{R_c}
  \newVar{\rserver}{R_s}

\newcommand{\vx}{\checkmark\kern-1.1ex\raisebox{.7ex}{\rotatebox[origin=c]{125}{--}}} 

\newcommand{\myalg}{2SYN\xspace}
\newcommand{\name}{\myalg}

\newcommand{\OPT}{\text{\textsc{opt}}}
\newcommand{\iopt}{i_{\OPT}}
\newcommand{\ALG}{\textsc{alg}\xspace}
\newcommand{\ALGi}{$\ALG_i$\xspace}
\newcommand{\ALGOPT}{$\ALG_{\iopt}$\xspace}

\begin{document}

\title{	
    \myalg: Congestion-Aware Multihoming
    
} 
\ifhyperref
    \newcommand{\aut}[2]{#1\texorpdfstring{$^{#2}$}{(#2)}}
\else
    \newcommand{\aut}[2]{#1$^{#2}$}
\fi

\author{
 \IEEEauthorblockN{
    \aut{Kfir Toledo}{1,2},
    \aut{Isaac Keslassy}{1}                
    }
\IEEEauthorblockA{
          $^1$ \textit{Technion} \quad 
          $^2$ \textit{IBM Research Haifa} 
    }   \BigReduceVSpace
    }

\OnlyTR{\markboth{Technical Report TR16-01, Technion, Israel}{}}


\maketitle


\thispagestyle{empty} 
\pagestyle{empty}     

\begin{abstract}
When sending flows to arbitrary destinations, current multihoming routers adopt simple congestion-oblivious mechanisms. Therefore, they cannot avoid congested paths.

In this paper, we introduce 2SYN, the first congestion-aware multihoming algorithm that works for any destination. We explain how it dynamically selects a preferred path for new connections, even given previously-unseen destinations. We further demonstrate that it can be easily implemented in Linux.
Finally, in a real-world experiment with either LTE or a wired link, we show how 2SYN dynamically adapts to the quality of the connection and outperforms alternative approaches. Thus, 2SYN helps companies better manage their networks by leveraging their multihoming capabilities.


\end{abstract}


\section{Introduction}\label{sec:introduction}

\begin{figure}[!t]
     \centering
            \includegraphics[width=0.8 \columnwidth]{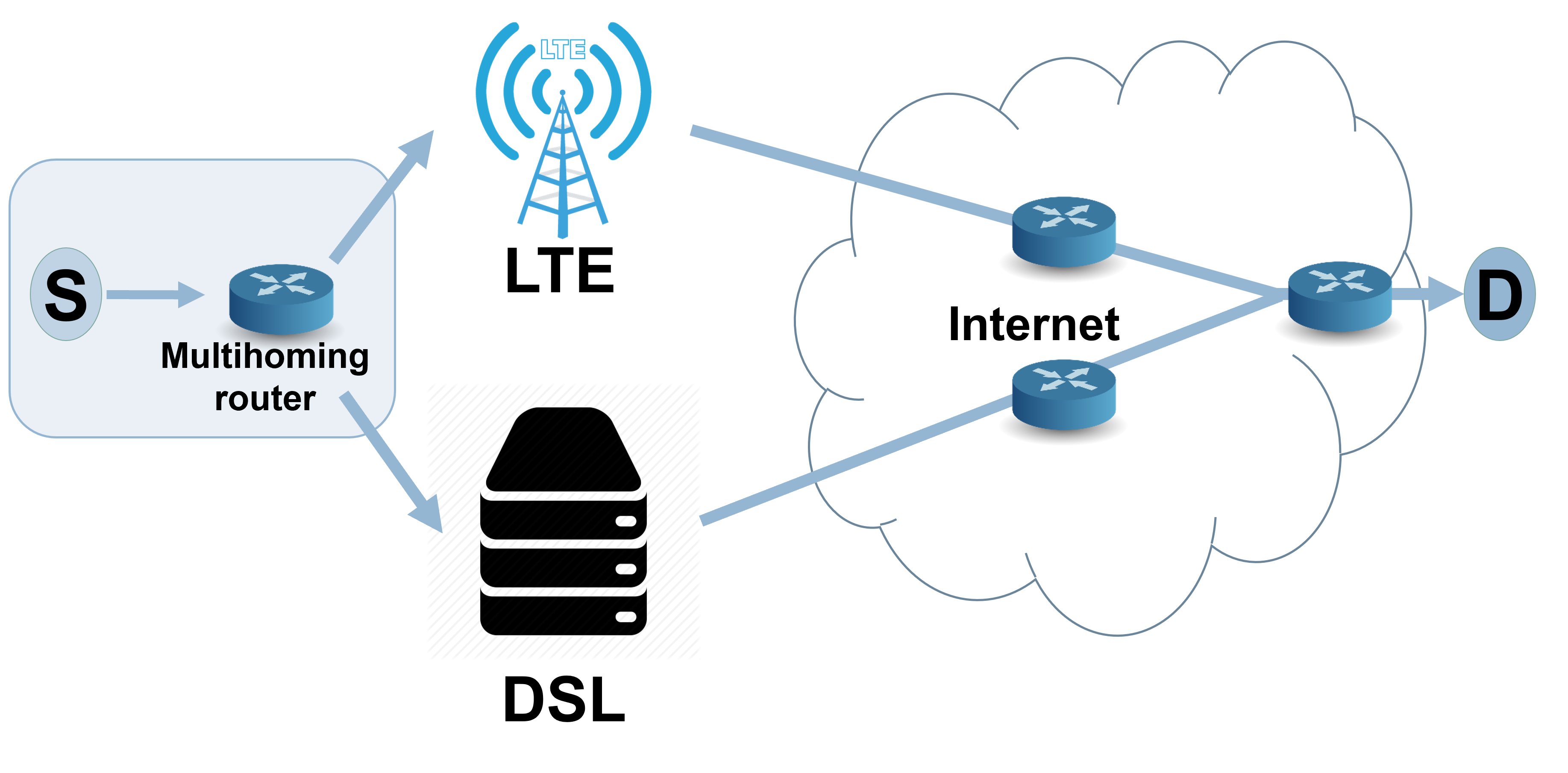}
         \caption{\textit{Multihoming overview}. \myalg runs in the multihoming router. 
         }
        \label{fig:system}
        \ReduceVSpaceA 
\end{figure}

To obtain a reliable Internet connection, companies are increasingly abandoning their expensive MPLS (Multiprotocol Label Switching) access, and using multihoming instead~\cite{mpls_vs_internet}. 
By correctly managing their multihomed network, 
companies can leverage several low-cost commercially-available internet-access links, such as LTE, DSL, cable, or fiber optics, to build a high-performance and high-availability WAN transport~\cite{wifi_lte_both}. As \fig{fig:system} illustrates, companies can use multihoming for any arbitrary website destination $D$, including 
public SaaS (Software as a Service) websites like Office365 or Workday. 

To decide which of the outgoing WAN links should be used, current multihoming routers adopt \textit{congestion-oblivious} 
algorithms: Either (1)~a static failover algorithm, 
which uses a preferred WAN link until it is disconnected, then uses a second preferred link, etc.; or (2)~a static load-balancing algorithm that load-balances flows across the links~\cite{draytek}. This load-balancing could be random or round-robin, uniform or weighted, but it is always congestion-oblivious. Thus, a large proportion of the connections are always at risk of suffering in a clogged or low-speed link, even when another high-speed link is uncongested.

In this paper, we focus on TCP flows, as they constitute the vast majority of corporate-oriented connections. When the destination $D$ belongs to the corporate network, the company can implement congestion-aware algorithms, \eg (1)~replace TCP by multi-path  protocols like MPTCP (Multipath TCP)~\cite{wifi_lte_both,0_rtt,amend2018robe}, or (2)~use SD-WAN (Software-Defined Wide Area Network) to establish tunnels to $D$ through each outgoing link, monitor the tunnels, and then choose the best link for current network conditions~\cite{sd_wan_survey,fajjari2017novel}.  
However, we want our algorithm to work for \textit{any} arbitrary destination $D$, \eg any SaaS website. Nothing guarantees that $D$ is configured to accept MPTCP, or that $D$ belongs to the corporate network. Therefore, these restricted algorithms do not apply to the general case.
\textit{The goal of this paper is to provide a general congestion-aware flow load-balancing algorithm given any arbitrary destination.}

\T{Related work.} As mentioned above, there are many solutions for MPTCP~\cite{wifi_lte_both,0_rtt,amend2018robe} and SD-WAN~\cite{sd_wan_survey,fajjari2017novel}. 
In particular, RobE~\cite{amend2018robe} sends SYNs over two parallel paths for an accelerated MPTCP handshake mechanism, although it does not later discard one of the SYN-ACKs. However, all these solutions assume that we have control over the destination $D$. Other works like CPR~\cite{landa2021staying} and the IETF architecture for transport services~\cite{brunstrom2022taps} deal with congestion-oblivious failover routing. 

\T{Contributions.} 
We introduce the 
\textit{\myalg } algorithm. \myalg leverages the TCP 3-way handshake protocol to determine the path with the best initial RTT. To do so, during the connection establishment to $D$, the router sends a SYN through each of the relevant router links. It then remembers the router link through which it obtains the first SYN-ACK, and keeps using it to route future packets while telling $D$ to disregard the other paths. We further implement \myalg in Linux and discuss its implementation tradeoffs.  
We also introduce alternative approaches based on ML (machine-learning) algorithms and explain why they appear inadequate.

We evaluate the performance of \myalg both in a lab testbed and in cross-continental transmissions. 
We show how it adapts to differing path propagation or queueing delays. We also show its resilience to a sudden bandwidth drop. Then, using long-haul transmissions and comparing LTE and {DSL links}, we confirm that \myalg tends to pick the best link, and adapts at connection time to artificial bandwidth drops or heavy congestion. We further compare it to alternative static, random, and ML-based approaches. We demonstrate how  \myalg can outperform them and significantly decrease the average FCT (Flow Completion Time) in a dynamic environment.

In summary, the main paper contributions are: (1)~The \myalg multihoming algorithm that dynamically avoids congested paths when establishing the connection to any destination $D$. (2) The \myalg implementation in Linux.  (3) The real-world experiments where \myalg outperforms the alternative static, random, and ML-based approaches. 

The open-source code of \name is available online~\cite{2syn_project}.

\section{The \myalg algorithm}\label{sec:alg}

\subsection{\myalg algorithm} \label{subsec:alg}

\begin{figure*}
 \centering
    \begin{subfigure}[b]{0.32\linewidth}
        \centering
        \includegraphics[width=1.02\columnwidth]{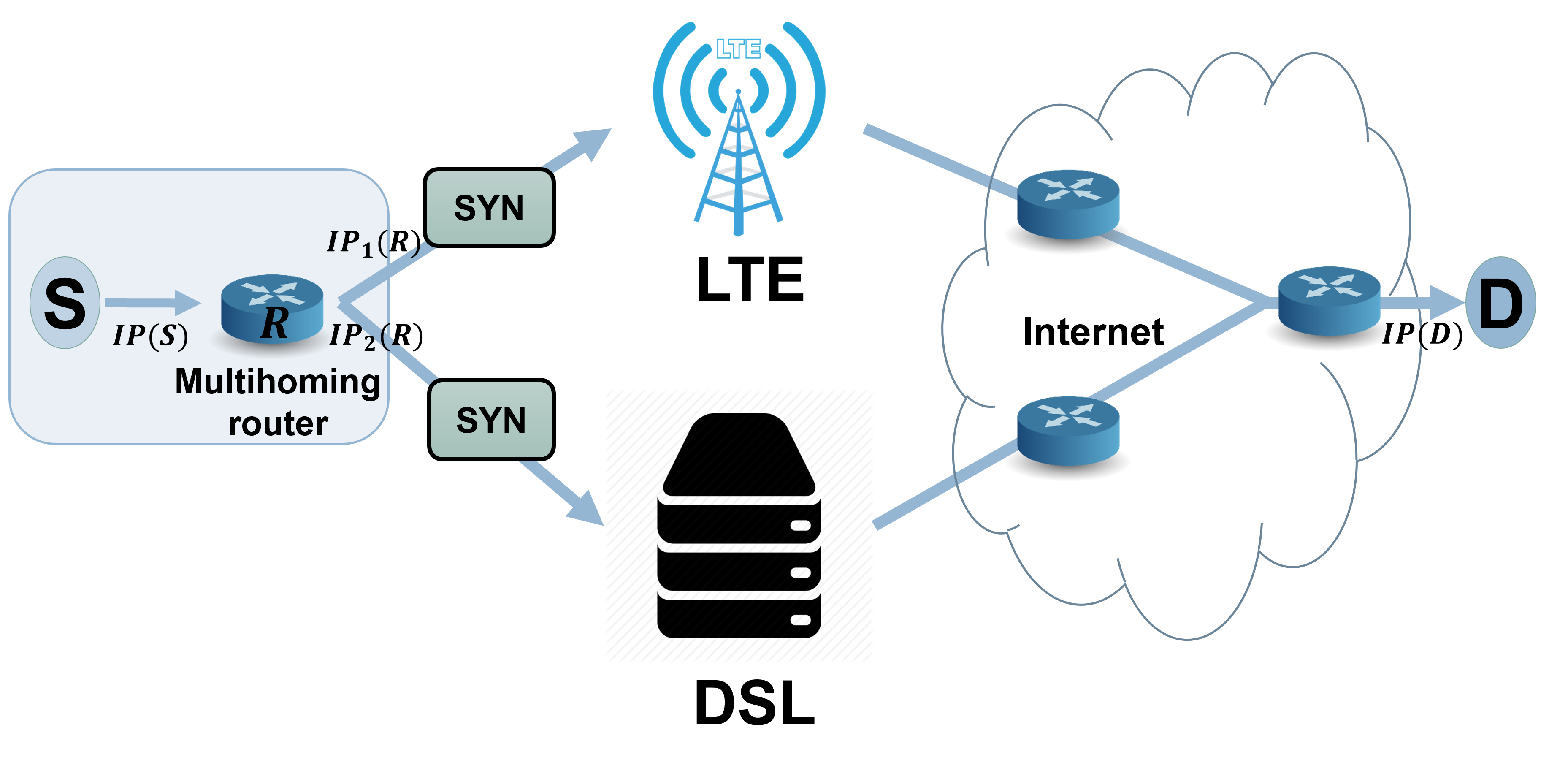}
        \caption{1st step: duplicate SYN}
        \label{subfig:1_step}
    \end{subfigure}
    \hfill
    \begin{subfigure}[b]{0.32\linewidth}
        \centering      
        \includegraphics[width=1.02\columnwidth]{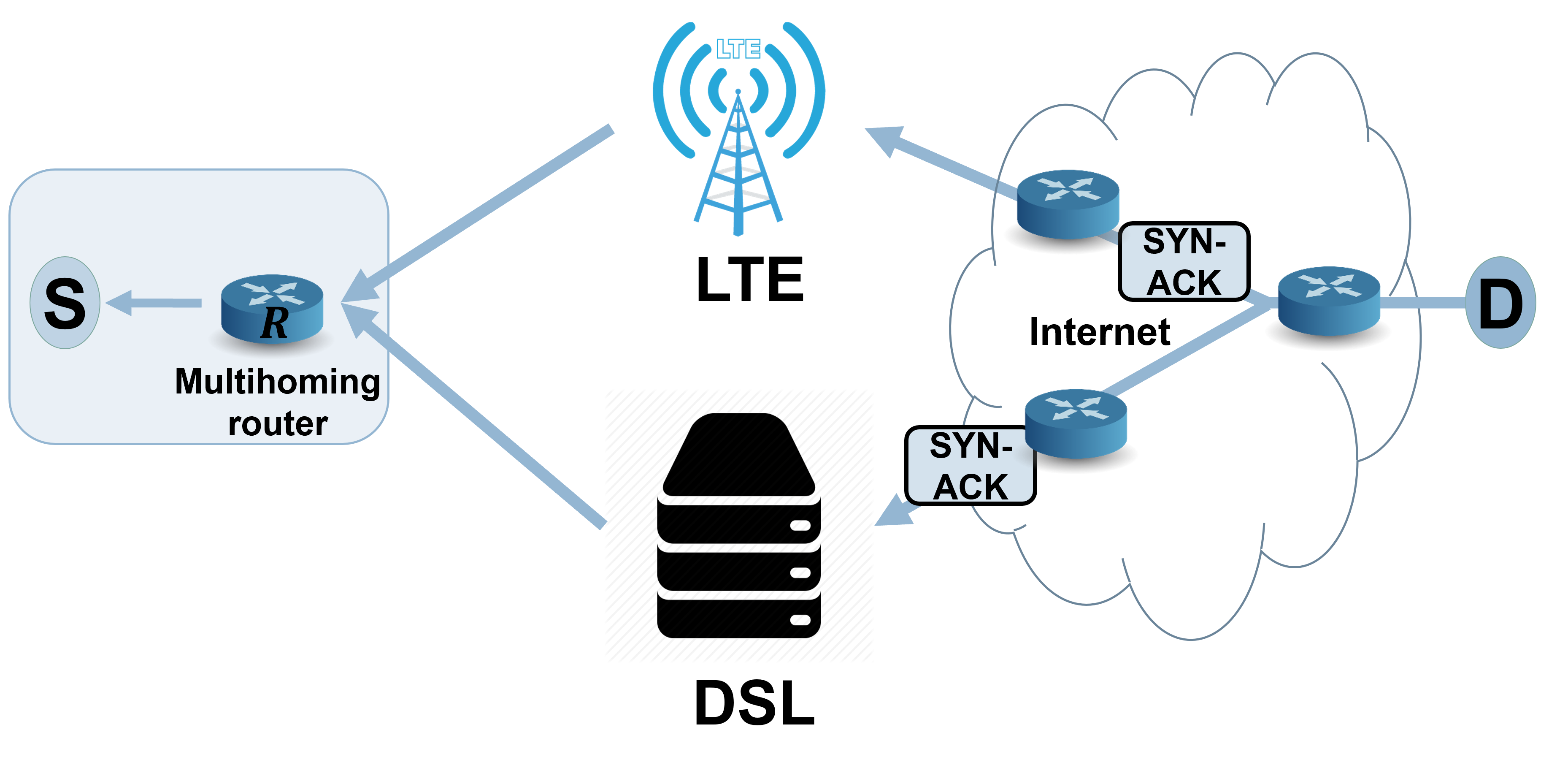}
             \caption{2nd step: wait for first SYN-ACK}
             \label{subfig:2_step}
     \end{subfigure}
     \hfill
     \begin{subfigure}[b]{0.32\linewidth}
        \centering      
        \includegraphics[width=1.02\columnwidth]{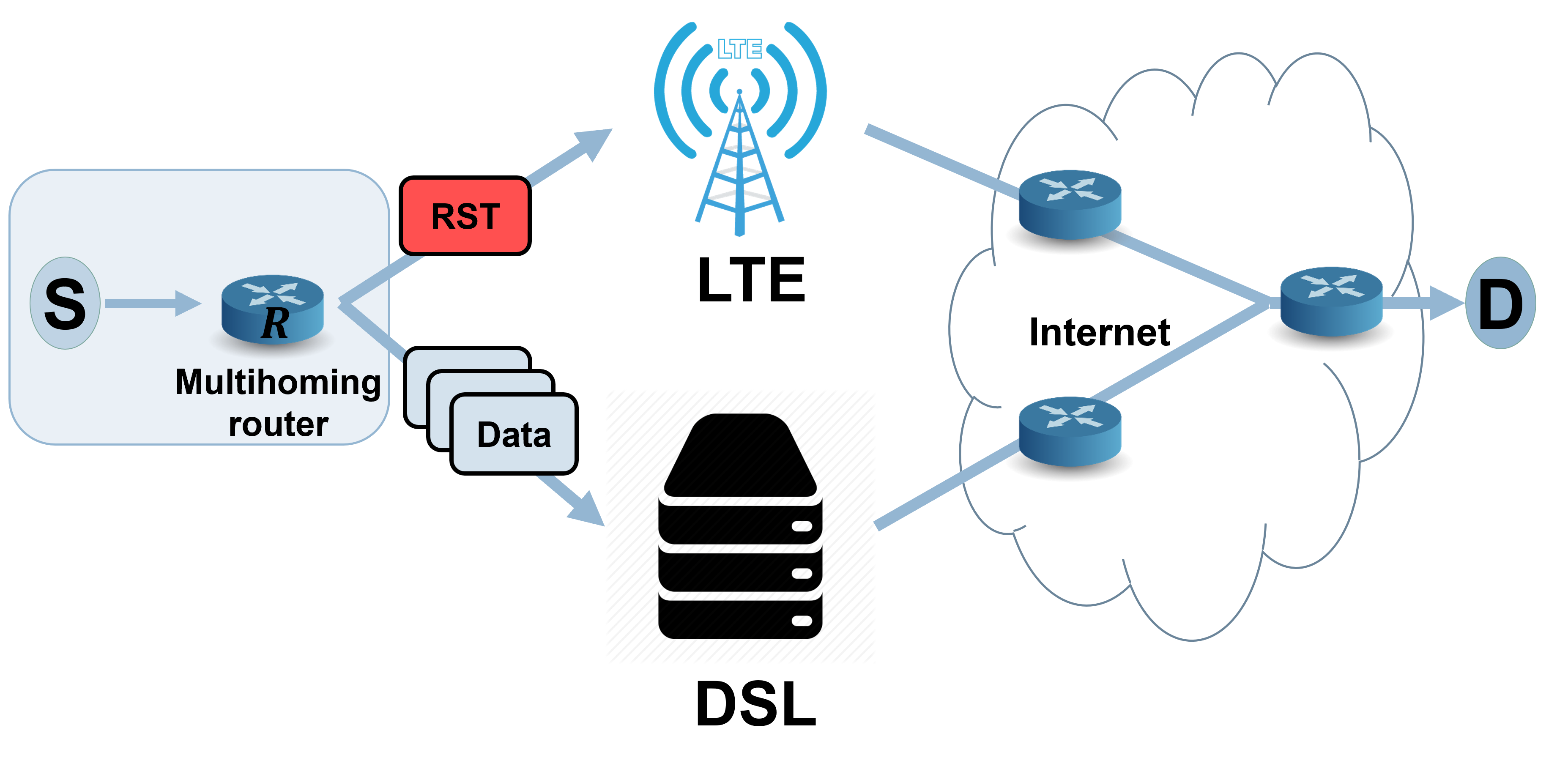}
             \caption{3rd step: pick path and cancel other paths}
             \label{subfig:3_step}
     \end{subfigure}
     \caption{Key steps of the \myalg algorithm.} 
    \label{fig:algorithm_steps}
    \ReduceVSpace
\end{figure*}

We introduce \textit{\myalg}, a congestion-aware load-balancing algorithm for a multihoming router $R$. The router $R$ is connected to $k$ different WAN links (\eg DSL or LTE), so each flow from $S$ to $D$ can take $k$ different possible paths.
We define a flow using its 5-tuple, and only consider TCP flows. 

\T{Key idea.}  \myalg  tries to choose the path with the shortest initial RTT by leveraging the TCP 3-way handshake protocol. When opening a TCP connection, it sends a SYN on each path, and only keeps the connection on the path through which it receives the first SYN-ACK, while canceling the other paths.



\T{Initialization.} We assume that \myalg manages an AF (Active Flow) table, which lists the current active flows that pass through the multihoming router.
AF is initially empty. Then, as illustrated in Fig.~\ref{fig:algorithm_steps}, \myalg performs the following steps:

\T{1. Duplicate SYN.} As shown in \fig{subfig:1_step}, 
for every new TCP flow, \ie for every new SYN that reaches the 
router $R$ from some source $S$ within the corporate branch, \myalg duplicates this SYN to the $k$ paths. 
As in the case without \name, since there are several router links with distinct ISPs, the \name router performs NAT (Network Address Translation) to replace the source IP address of each SYN by the IP address of the corresponding router link. For example, in \fig{subfig:1_step}, the source IP address of the top SYN is $IP_1(R)$ instead of $IP(S)$. 


\T{2. Wait for the first SYN-ACK.} Later, the destination $D$ receives the $k$ SYNs. Since each SYN has a different source IP, $D$ treats them as $k$ independent new-flow requests. Therefore, as \fig{subfig:2_step} illustrates, it sends a SYN-ACK for each of these flows along their respective paths. 

\T{3. Pick a path and cancel the other paths.} When the first SYN-ACK reaches the router, \myalg chooses the corresponding link as the preferred link for this flow. 
\new{(1)~As shown in \fig{subfig:3_step}, \myalg sends RST messages to $D$ via the other $k-1$ links, causing $D$ to cancel them.
(2)~\myalg updates the routing table to route the remaining packets through the chosen link, and adds the flow to the AF table.}

\T{4. Flow completion.} \myalg then waits for the flow to complete in its selected path (starting with a FIN or RST flag from $S$ or $D$). When the flow completes (or a  timeout expires), \myalg removes it from the AF table and deletes its rule from the routing table.

While we described the algorithm for the general case of $k$ paths,  
we focus on 
$k=2$ in the remainder for simplicity. 

\subsection{\myalg implementation in Linux} 
\label{subsec:impl}

A recent report unpacked 89\% of the measured home routers and found that all of them were based on Linux~\cite{vom2022home}.  Given this Linux ubiquity, we develop a proof-of-concept implementation of \myalg in Linux~\cite{2syn_project}.
For any packet that reaches the router, \myalg uses iptables to check whether it is a SYN packet for a new TCP flow from $S$ to $D$. Specifically, given any packet with a SYN flag, it checks whether its flow satisfies the following four conditions: 
(1)~It does not already have a specific rule in the Linux routing table, 
(2)~it is a TCP flow, 
(3)~it comes from a pre-defined router interface corresponding to the corporate branch, and 
(4)~it is not destined to a pre-defined corporate gateway router through an existing tunnel. 
If all conditions are satisfied, \myalg uses iptables to automatically duplicate the SYN packet to all outgoing router interfaces and update the corresponding source IP. 
Else, it is routed with the standard Linux route rules. 

We use the Python Scapy API~\cite{scapy} to monitor traffic. 
To accelerate packet processing, Scapy API filters packets using BPF (Berkeley Packet Filter)~\cite{bpf} and only monitors packets with set SYN, SYN-ACK, FIN and RST flags.
When a SYN arrives, \myalg knows that it needs to wait for its first SYN-ACK. Then, when the first SYN-ACK arrives, it adds the flow to the AF table and the Linux routing table. 
Finally, with a FIN or RST, \myalg removes the flow from both the AF table and the Linux routing table.

In addition, we implemented the option to 
redefine a flow as a (source IP, dest. IP) pair, neglecting 
ports and protocol. This allows sharing routing decisions among flows with the same (source IP, dest. IP) pair, reducing SYN duplication rates at the cost of a more complex table management. 

\subsection{Overhead}
\T{Router overhead.} 
To minimize the computation overhead, our \myalg implementation delegates to the Linux router both the path routing computation and the SYN duplication operation, both with negligible time. The only non-negligible computation overhead is in updating the Linux routing table, either to add a new flow or to remove a completed flow. Using our simple Python Scapy implementation, it takes some 4~ms on average.
In addition, adding and removing entries in the AF table is done in an $O(1)$ negligible time by using the hash-table data structure. 
Using BPF, the \myalg algorithm processes only packets with the SYN, SYN-ACK, FIN, and RST flags, so even for high flow rates, the processing time is negligible compared to the update time of the Linux routing table. We checked with mpstat~\cite{mpstat} the CPU usage overhead when processing a sudden burst of $100$ parallel SYN arrivals. At peak time, the  CPU overhead is  $3\%$.

\T{Traffic overhead.} For each new TCP flow,  \myalg  creates one additional SYN (\ie two SYNs instead of one), one SYN-ACK, and one RST (for the non-selected path). This appears to be negligible in practice.

\T{Destination overhead.} 
For each new flow, we send an additional SYN to $D$, and cancel it with a RST after one RTT. Thus, we roughly increase the number of half-open connections at $D$ by $RTT/FCT_D,$ where $FCT_D$ denotes the average FCT as seen by $D$. 
Assuming a median $FCT_D$ of about 100~RTTs (the median flow lasts about 6~sec.~\cite{bauer2021evolution}), the overhead is around 1\%, which seems reasonable.




\T{NAT overhead.}  
In \name, the NAT overhead is negligible: we translate 2 SYN/SYN-ACK messages per flow instead of 1, and proceed with other packets without change.


\TR{
\subsection{\name optimality} 

In some cases, \name can be proved to be optimal. Denote by 
\ALGi 
the algorithm that always sends packets through a path going through multihoming-router link $i$, which we call path $i$, 
and neglect processing time. \fig{fig:2syn_timig_plot} 
illustrates a case where path $1$ is the path with the shortest RTT for the SYN/SYN-ACK. It provides some intuition on why \name will have the same connection establishment time as $\ALG_1$, and later the same FCT.
Therefore, whenever this shortest initial RTT also leads to the best FCT, \name is \textit{optimal}, in the sense that it has the same FCT as the optimal \ALGi.

We prove this below in two simple cases. For simplicity, we rely in both cases on  the following assumptions: Only $S$ sends traffic and not $D$, all of its packets are ready to be sent,  there are no link losses, processing times and queueing times of non-data packets like SYNs and ACKs are negligible, all the multihoming-router egress links have the same capacity, 
and all flows use a standard TCP like New Reno, with or without the TCP-SACK option, but with sufficiently large timeouts and receive windows.
The proofs are in the Appendix. One of the interesting aspects of these results is that even though they are fairly intuitive, the complexity of the TCP mechanisms leads to a longer list of assumptions and longer proofs than would be expected.

\TR{
\begin{figure}
 \centering
    \begin{subfigure}[b]{0.45\linewidth}
        \centering
        \includegraphics[width=\columnwidth]{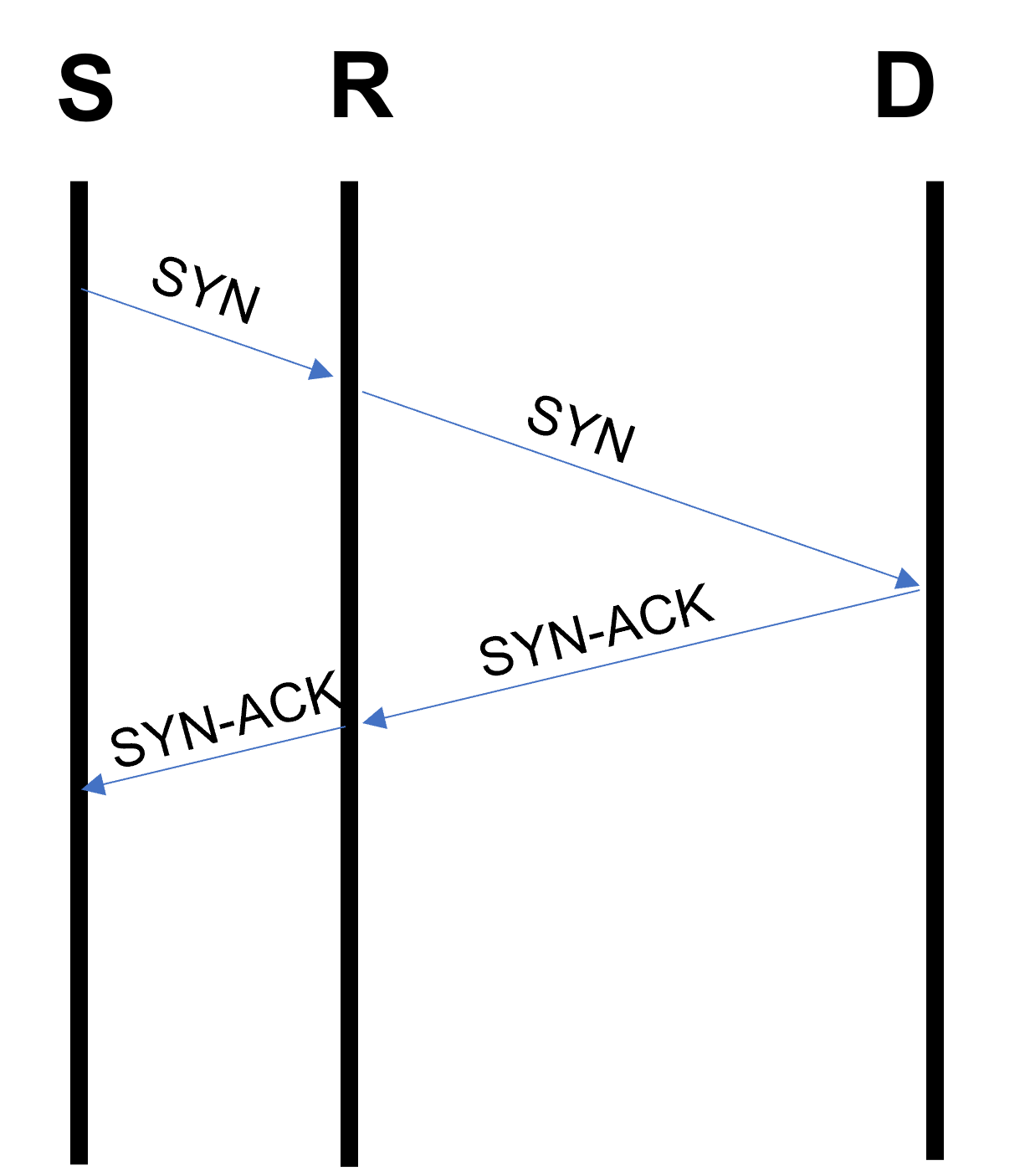}
        \caption{Without the \name algorithm}
        \label{subfig:without_2syn}
    \end{subfigure}
    \hfill
    \begin{subfigure}[b]{0.45\linewidth}
        \centering      
        \includegraphics[width=\columnwidth]{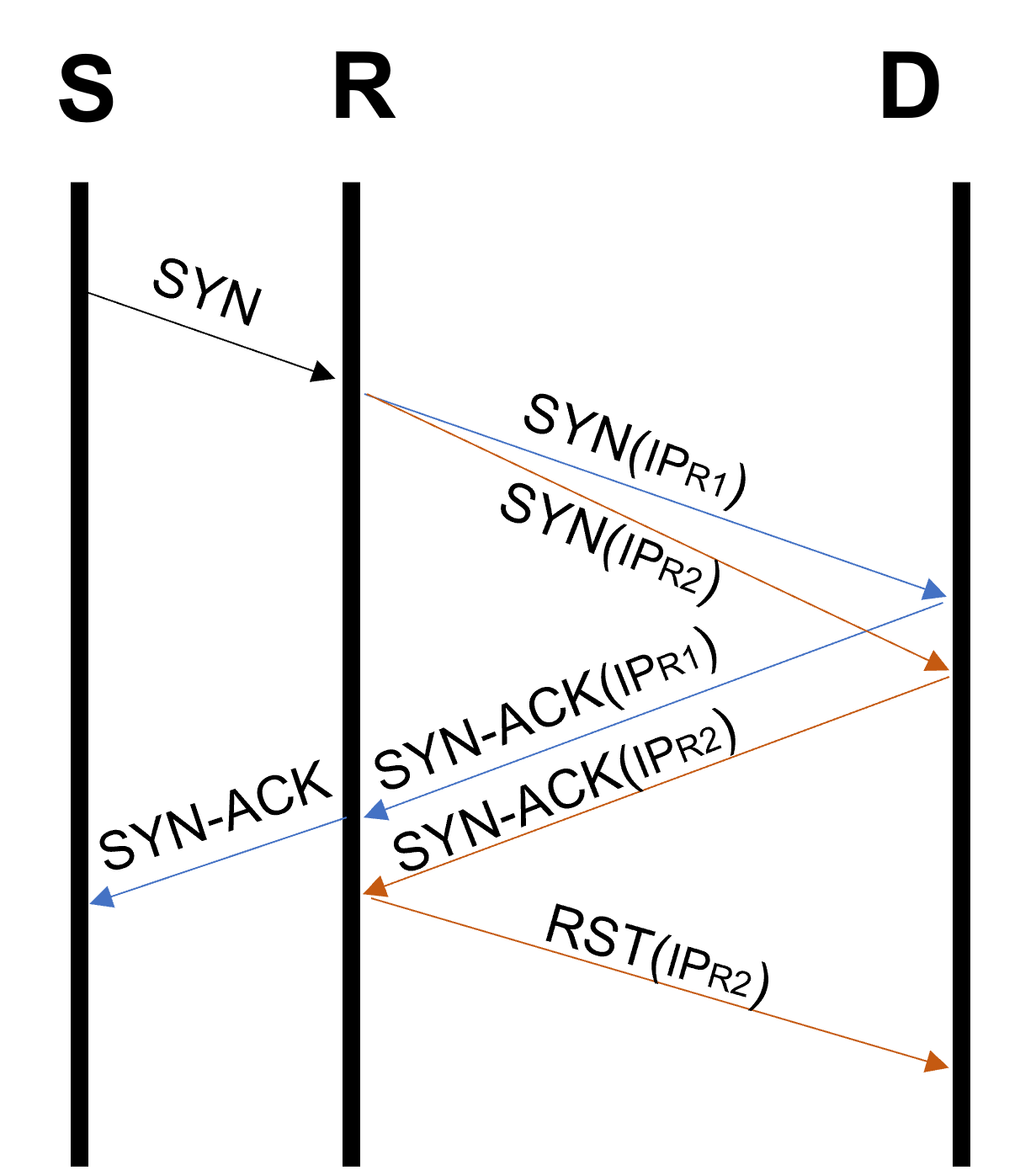}
             \caption{With the \name algorithm}
             \label{subfig:with_2syn}
     \end{subfigure}
     \caption{Timing diagram of the \myalg algorithm.}
    \label{fig:2syn_timig_plot}
\end{figure}
}
\begin{theorem}[Lower RTT]\label{thm:general}
Assume that 
the RTT for \textit{any} (packet, ACK) pair is strictly lower over path $\iopt$ than over any other path $i$.
Further assume that packets and ACKs experience no reordering on any path. Then:
\begin{enumerate}[label=(\roman*)]
    \item \ALGOPT is optimal, \ie it achieves the lowest FCT  among all algorithms \ALGi.
    \item \name  is optimal, \ie it achieves the same lowest FCT as  \ALGOPT. 
\end{enumerate}
\end{theorem}

\begin{corollary}[Lower Propagation Delay]\label{cor:1}
Assume that all paths have arbitrary propagation delays, and the round-trip propagation time in path $\iopt$ is strictly lower than in other paths $i$. Further assume that all paths have the same bottleneck capacity in the forward (resp., backward) direction and negligible queueing delay. Then \ALGOPT and  \name are optimal.
\end{corollary}
} 

\section{Alternative approach}\label{subsec:alternative_approaches}

We suggest an alternative lightweight ML-based approach, assimilating the problem to a  Multi-Armed Bandits (MAB) problem. In the MAB setting, $k$ slot machines offer random rewards, drawn from $k$ unknown distributions. The goal is to maximize the cumulative reward by selecting machines wisely. Similarly, in our routing problem, the $k$ paths represent slot machines, offering different FCTs. We aim to choose the path with the lowest FCT (highest reward), balancing exploitation (choosing paths with historically low FCTs) and exploration (sampling less-traveled paths). We consider three standard MAB algorithms: \textit{$\epsilon$--greedy}, \textit{Upper Count Bound (UCB)}, and \textit{Thompson sampling} (see Burtini et al.~\cite{burtini2015survey}). 
For instance, for some small $\epsilon>0$, $\epsilon$--greedy chooses the historically-optimal path with a probability of $1-\epsilon$, and explores other paths uniformly at random otherwise. 
These MAB algorithms are lightweight, do not require dedicated hardware (\eg GPUs), and do not need pre-training for each possible $D$.

\T{Limitations.} 
\new{MAB algorithms} suffer from several limitations. First, the analogy between slot machines and paths is misleading: Choosing a path reduces its future performance due to the increased congestion. This differs from the usual MAB setting.
Second, as we illustrate in the experiments (\se{sec:evaluation}), the path properties can vary with time in a non-stationary way, while MAB algorithms are not very reactive, which hurts future connections. 
Lastly, we would need to maintain a large database of flows with their past performances for each $D$. 

\section{Experiments}\label{sec:evaluation}



\subsection{Lab experiments}\label{subsec:lab_exp}

We first evaluate \name in a lab testbed, then in real-world conditions.
In the following lab testbed experiments, we start by studying the impact of 
propagation delay and queueing delay.
Then, we study the resilience to bandwidth drops.

\T{Setting.}
We 
use five 
Linux servers with Ubuntu 20.04. As in \fig{fig:system}, 
they emulate the source $S$ (iPerf3 client), the multihoming router, the two paths using two servers
(path routers), and the destination $D$ (iPerf3 server).

\U{Tools.} In all experiments, we use the iPerf3 (version 3.9)~\cite{iperf3_link} network measurement tool to run the tests, download/upload a file and measure the average FCT. We also use the NetEm~\cite{netem_tool} network emulation tool to control the RTT and bandwidth of each path. The buffer size of each path router of bandwidth $BW$ is set to $RTT \cdot BW$. 


\U{Algorithms.}  We compare four different algorithms.
The first always chooses the first path, and the second chooses the second path. The third chooses the path randomly with equal probabilities. The last algorithm is \myalg. 

\begin{figure}
 \centering
    \hspace*{\fill}
    \begin{subfigure}[b]{0.48\columnwidth}
    \centering             
    \includegraphics[width=1.05\columnwidth]{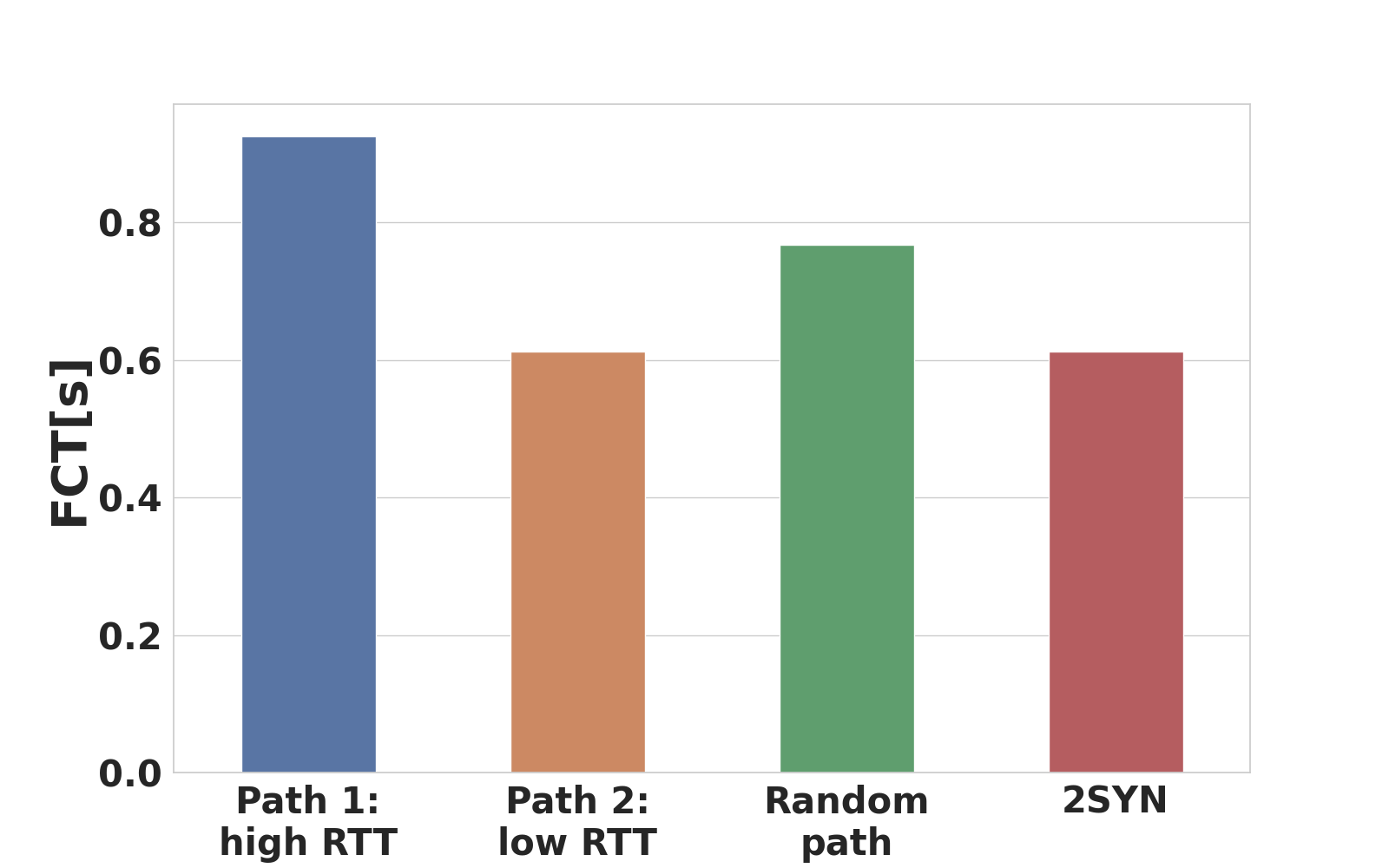}
         \caption{Different propagation delays}
         \label{subfig:diff_rtt_1m}
     \end{subfigure}
     \begin{subfigure}[b]{0.48\columnwidth}
        \centering
        \includegraphics[width=1.05\columnwidth]{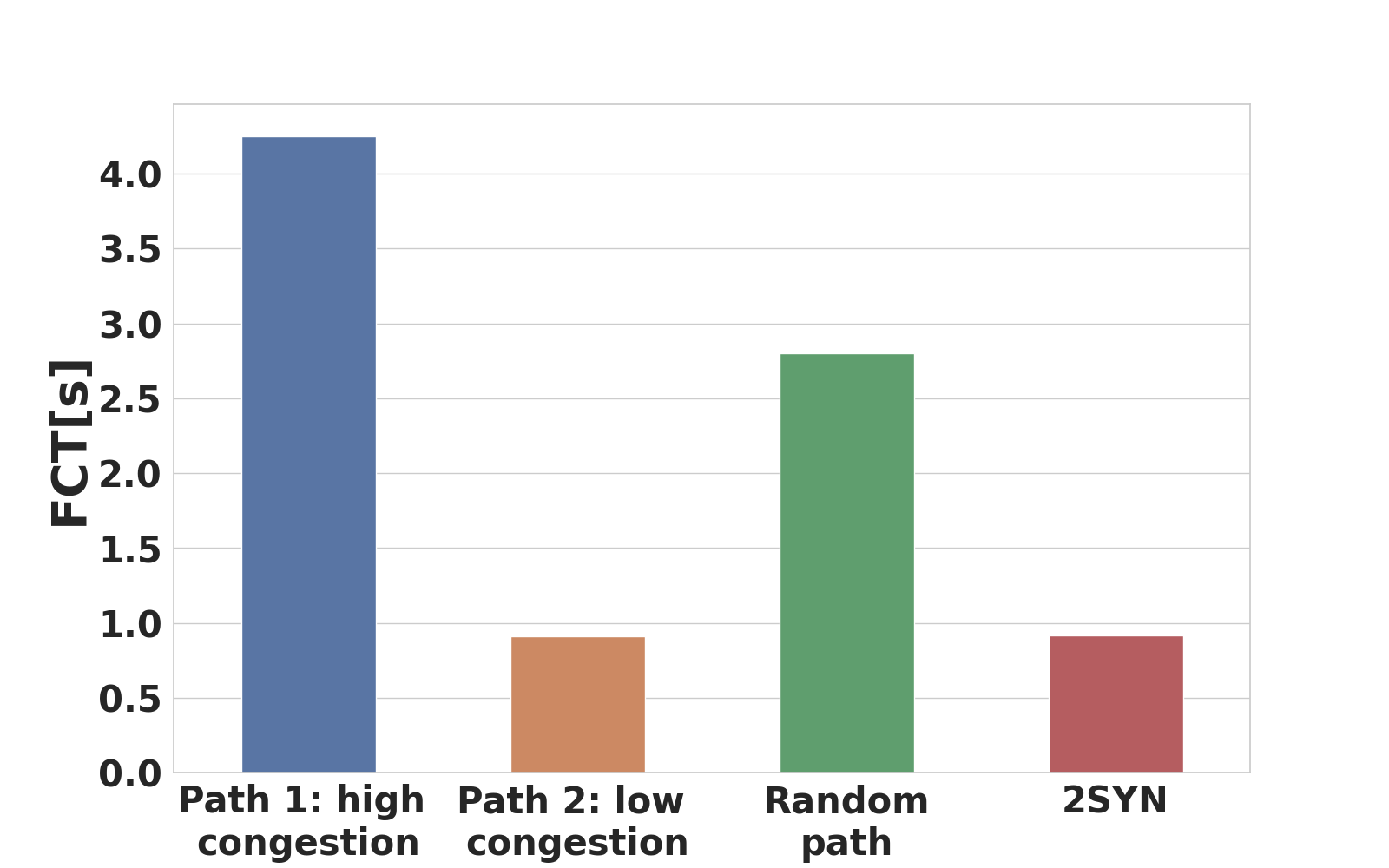}
        \caption{Different queueing delays}
        \label{subfig:diff_queue_1m}
     \end{subfigure}
    \hspace*{\fill}
     \caption{\textit{Paths with different delays.} Average FCT in
     a 1-MB file download using (a)~different propagation delays, and (b)~different queueing delays. 
     }
    \label{fig:diff_rtt}
    \ReduceVSpace 
\end{figure}

\T{Impact of propagation delay.} 
For each WAN path, we set the same bandwidth of $300$\,Mbps, 
but a different propagation delay: 
$120$\,ms for the first path and $80$\,ms for the second. Then, we measure the FCT to download a 1\,MB file from $D$ (a 100\,MB file download yielded similar results).
We repeat the experiment 20 times and measure the average download time. \fig{subfig:diff_rtt_1m} illustrates the results. \myalg  always chooses the best path with the lowest propagation delay, providing FCTs that are comparable to those of path 2. 

\T{Impact of  queueing delay.} 
We now set both paths with the same bandwidth of $300$\,Mbps and RTT of $120$\,ms, but add background TCP traffic (five flows with a limit of 100\,Mbps per flow) to path 1. 
We measure the average download time for a 1\,MB file, using 20 experiments. \fig{subfig:diff_queue_1m} shows that \myalg manages to choose the path with less congestion, again providing FCTs  comparable to those of path 2.

\begin{figure}
 \centering
    \begin{subfigure}[b]{0.48\columnwidth} 
        \centering
        \includegraphics[width=0.97\columnwidth]{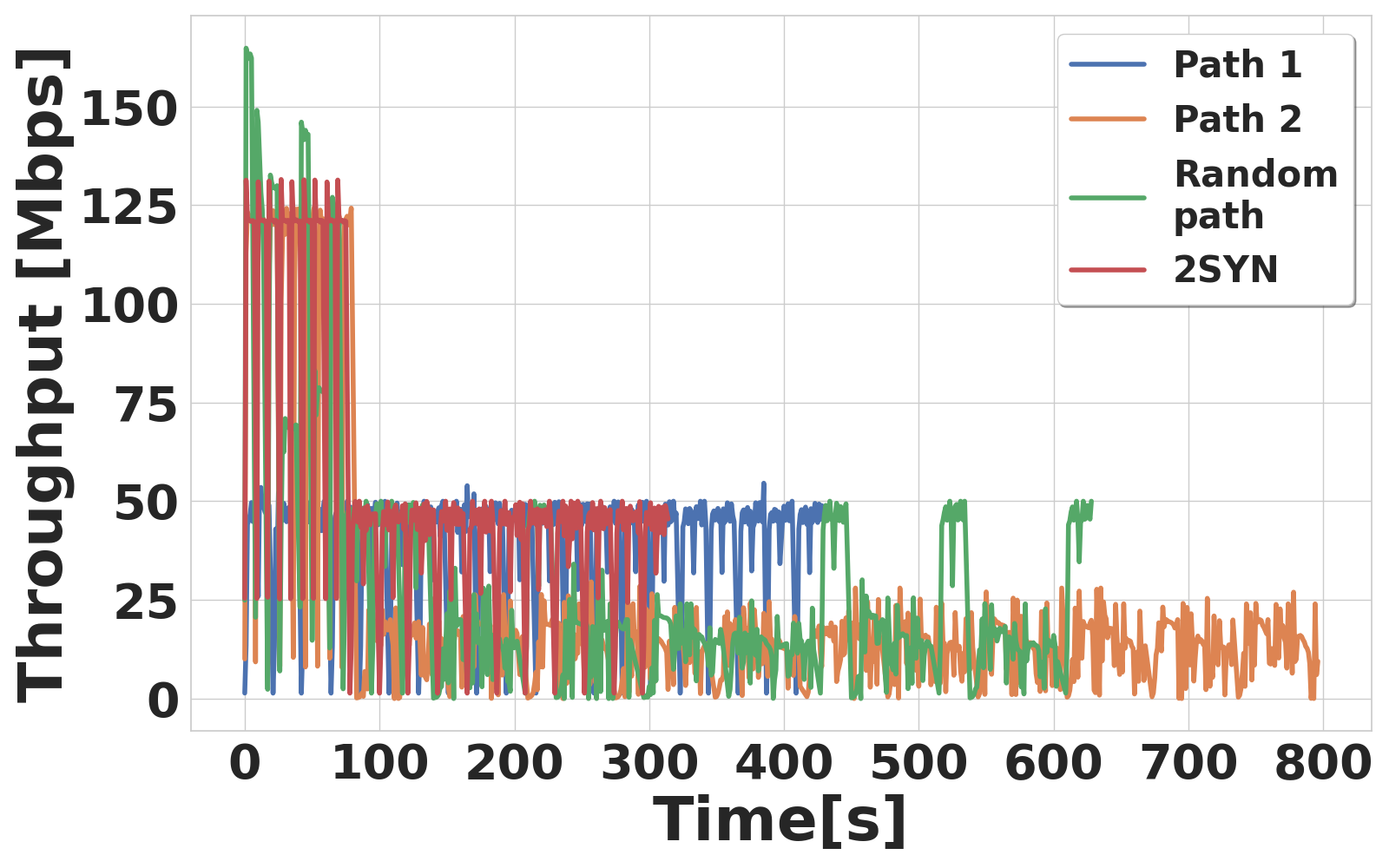}
        \caption{Throughput over time}
        \label{subfig:diff_bw_terminate_th}
    \end{subfigure}
    \begin{subfigure}[b]{0.48\columnwidth}  
        \centering      
        \includegraphics[width=1.05\columnwidth]{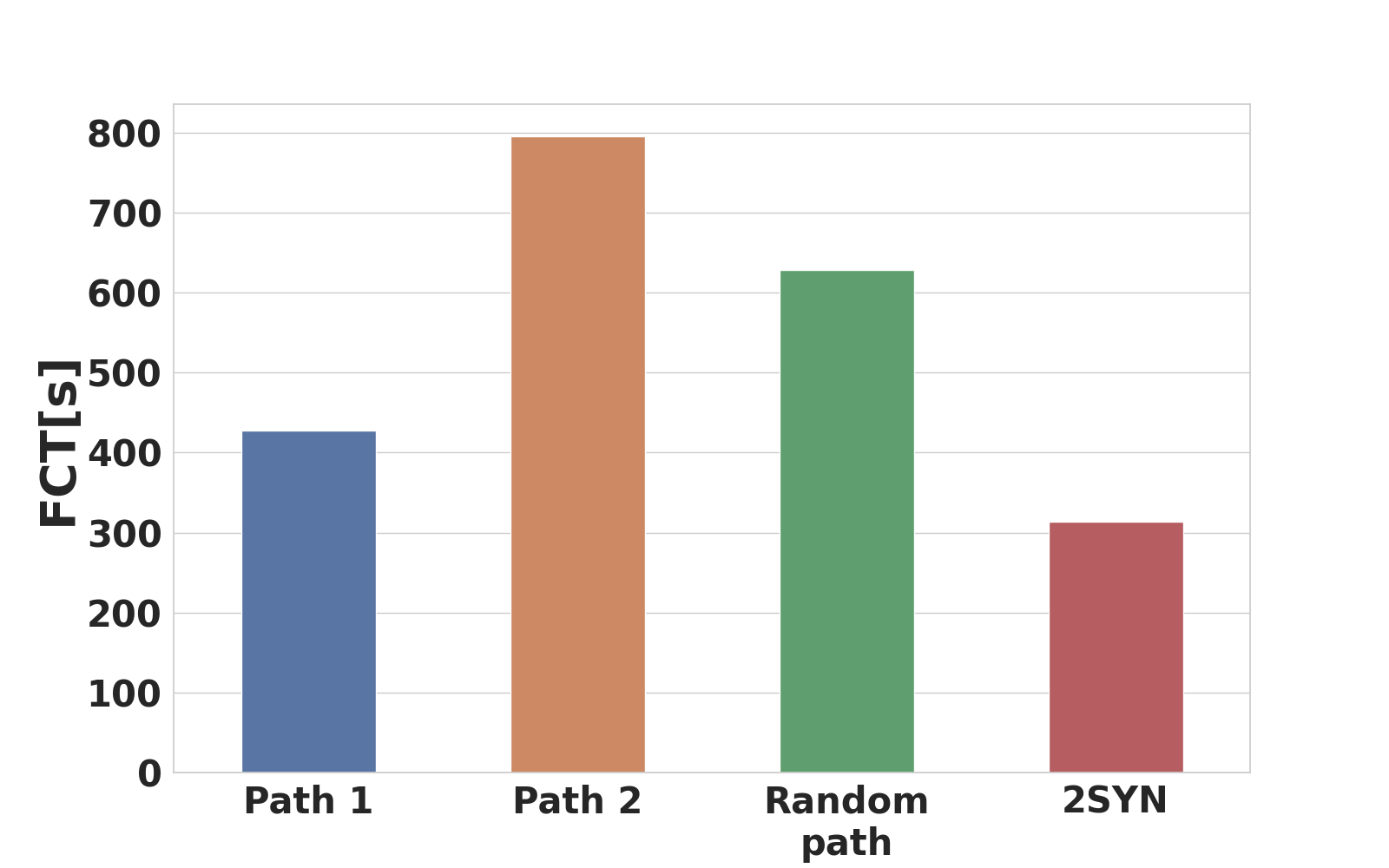}
             \caption{FCT}
             \label{subfig:diff_bwe_fct}
     \end{subfigure}
     \caption{\textit{Resilience to bandwidth drop.} Path 1 has a constant bandwidth of 100 Mbps,  while 
     that of path 2 suddenly drops from  300\,Mbps to 30\,Mbps.  
     }
    \label{fig:diff_bw_terminate}
    \ReduceVSpace 
\end{figure}

\T{Resilience to bandwidth drop.} In \fig{fig:diff_bw_terminate}, we 
download 100-MB files 20 times in a row. We reduce the bandwidth of path~2 from 300\,Mbps to 30\,Mbps after downloading 40\% of the files,
while 
that of path 1 remains constant at 100\,Mbps. 
\fig{subfig:diff_bw_terminate_th} shows how the throughput varies with time for all algorithms. The throughput oscillates due to TCP's slow start for each new flow. 
\new{Significantly, \myalg chooses the high-bandwidth path for the first connections, while after the bandwidth drop, it adopts the other path for new connections. \fig{subfig:diff_bwe_fct} shows that it achieves a lower FCT than 
any constant-path algorithm.}

\subsection{LTE \vs DSL experiments}

\T{Setting.} We download or upload files to public SpeedTest servers in England~\cite{speed_test_public_server}, with our source $S$ at the Technion, Israel. 
We connect our source router to (1)~a standard cell phone with a 4G LTE connection; and (2)~a wired internet link that is throttled to $100$\,Mbps for download and $10$\,Mbps for upload, based both on the median offered DSL package across all providers and on the median measured
broadband bandwidth~\cite{israel_okla}. 
We run two workloads: (1)~sending a \textit{large file} of $30$\,MB  for $20$ times, or (2)~sending \textit{web-search} application traffic with a distribution of $62\%$ for small files ($<100$\,KB), $18\%$ for medium files ($100$\,KB$-1$\,MB), and $20\%$ for large files ($>1$\,MB)~\cite{web_search_1}. 
We test three scenarios: one without issues; one with a bandwidth drop in the middle of transmission, modeling a sudden path router or ISP problem; and one with a sudden congestion increase caused by additional users in the middle of transmission. We get 2 transfer modes $\times$ 2 workloads  $\times$ 3 scenarios, \ie 12 experiments. We find that all results are similar, and show 4 typical ones.
    
\begin{figure}
 \centering
    \begin{subfigure}[b]{0.48\columnwidth}
    \centering             
    \includegraphics[width=1.05\columnwidth]{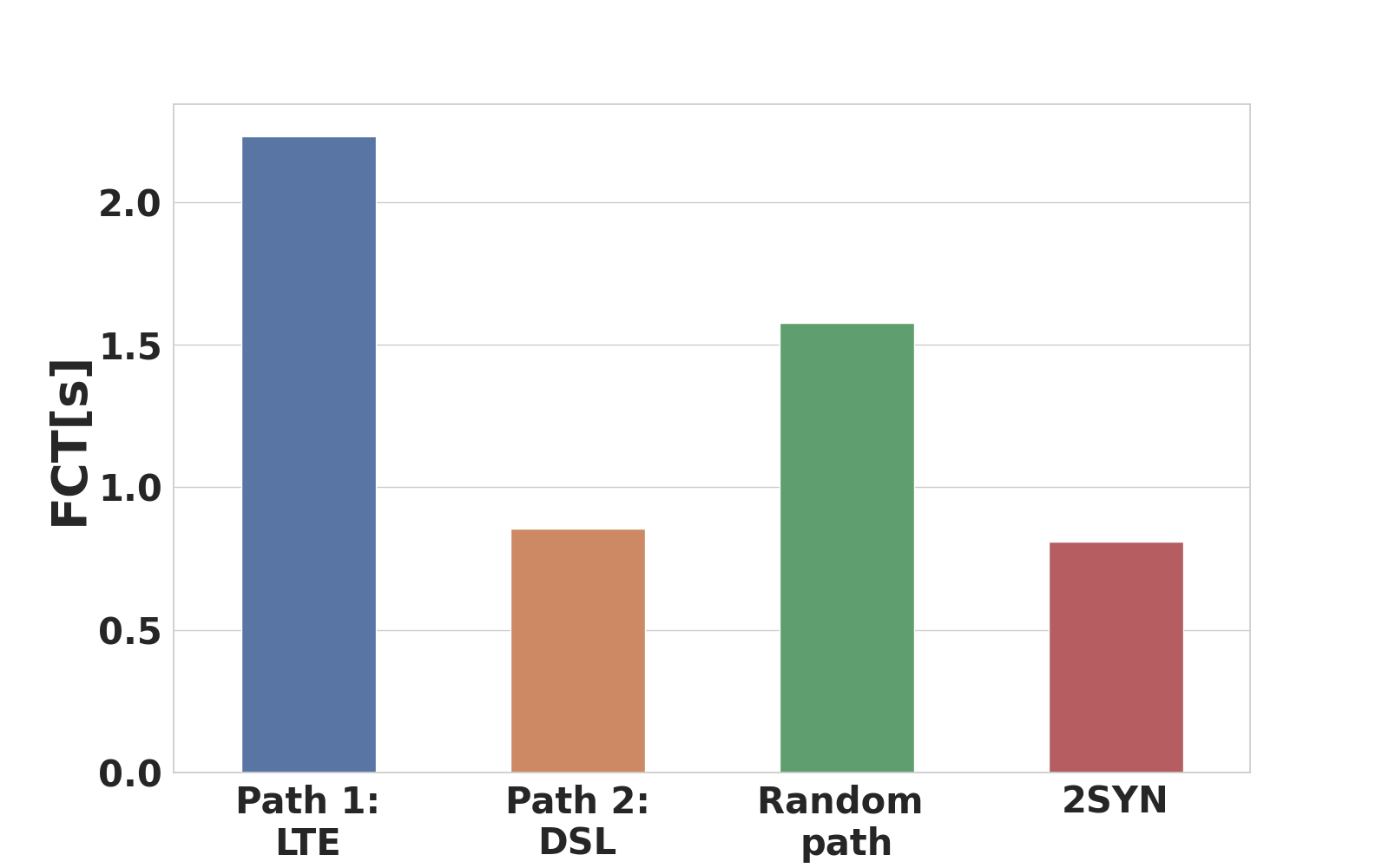}
         \caption{Links without issues}
         \label{subfig:simple_test_Download}
     \end{subfigure}
     \begin{subfigure}[b]{0.48\columnwidth}
        \includegraphics[width=1.05\columnwidth]{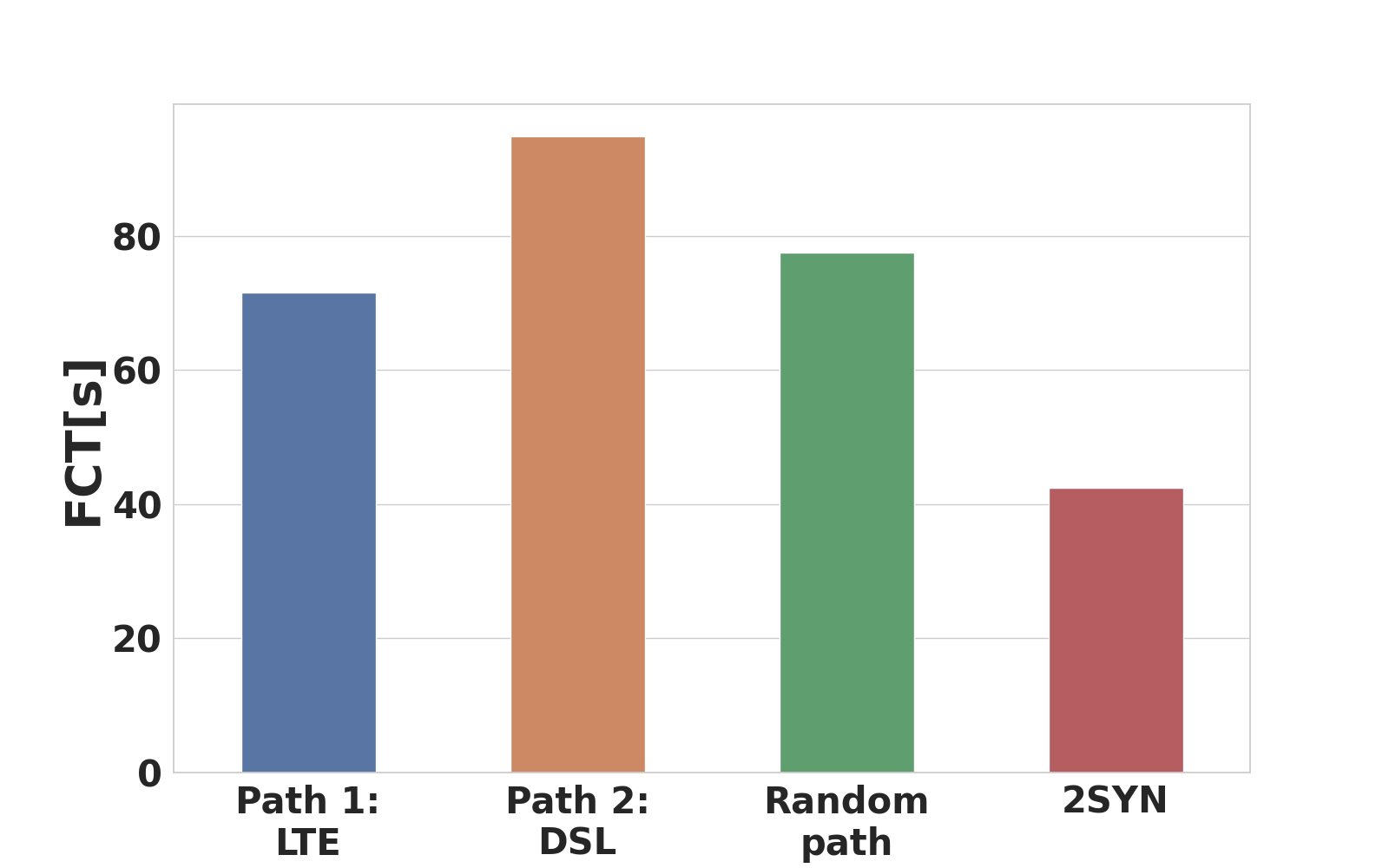}
        \caption{Links with bandwidth drop}
        \label{subfig:lte_wired_dropout_web_download}
     \end{subfigure}
     \caption{\textit{LTE (path 1) \vs DSL (path 2).} Average FCT for the download of web-search files. 
     }
    \label{fig:lte_wired_simple_test}
    \ReduceVSpace 
\end{figure}

\T{Links without issues.} 
\fig{subfig:simple_test_Download} shows how in all cases, \myalg keeps choosing the better DSL link as expected.

\T{Links with bandwidth drop.} \fig{subfig:lte_wired_dropout_web_download} illustrates a case where we arbitrarily drop the DSL download bandwidth to $5$\,Mbps after $40$\% of the files are sent.
\myalg recognizes when the DSL bandwidth drops and switches to the LTE path for new flows. Its performance is better than sticking to any path.

\begin{figure}
 \centering
    \begin{subfigure}[b]{0.48\columnwidth}
    \centering             
    \includegraphics[width=1.05\columnwidth]{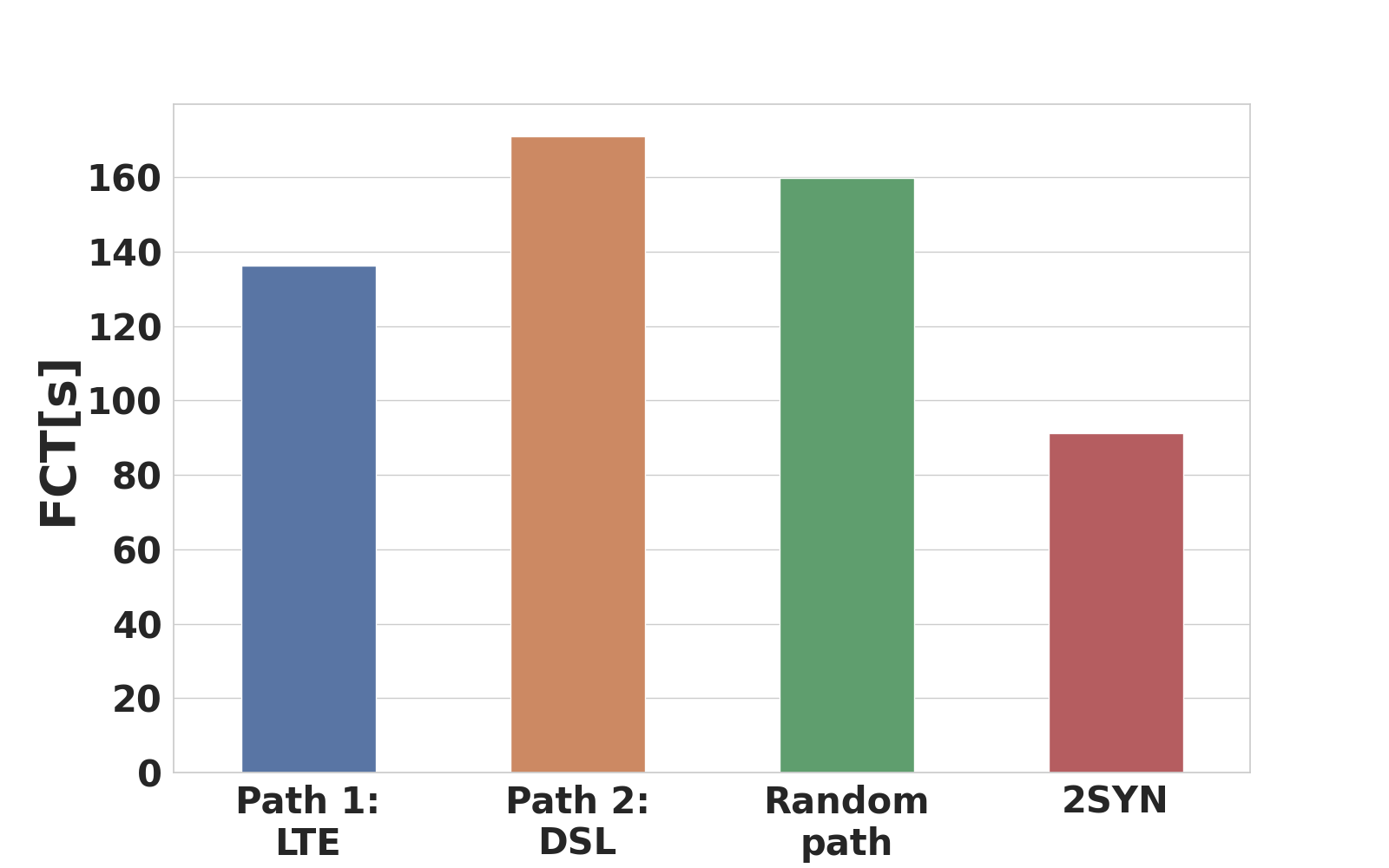}
         \caption{Web search: download link}
         \label{subfig:subfig:lte_congst_web_Download}
     \end{subfigure}
     \begin{subfigure}[b]{0.48\columnwidth}
        \includegraphics[width=1.05\columnwidth]{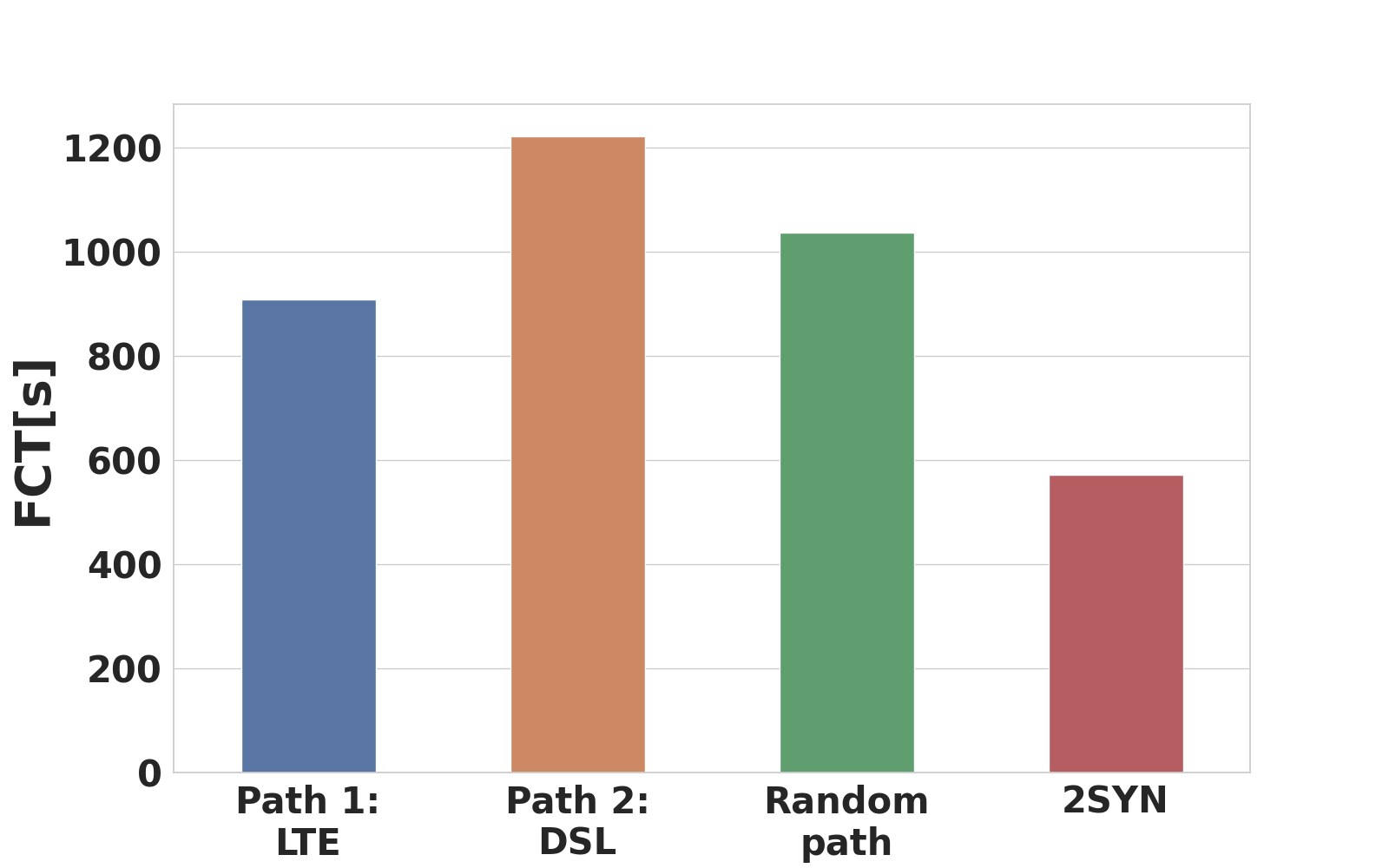}
        \caption{30\,MB file: upload link}
        \label{subfig:subfig:lte_congst_30m_upload_link}
     \end{subfigure}
     \caption{\textit{LTE (path 1) \vs DSL (path 2) with sudden congestion}, due to other users on the DSL link. 
     }
    \label{fig:lte_heavy_congs}
    \ReduceVSpace
\end{figure}

\T{Links with heavy congestion.} \fig{fig:lte_heavy_congs} shows the impact of setting a significant congestion after $40$\% of the files are sent, by adding ten TCP flows to the DSL link to model congestion from other users.
\myalg quickly abandons the congested DSL link, thus yielding the best performance.

\subsection{Why not use ML?}
We compare \myalg to the MAB ML algorithms (\se{subsec:alternative_approaches}) in the lab testbed. 
We set the bandwidth to $200$\,Mbps for the first path and $300$\,Mbps for the second. We  send traffic on five (source, dest.) pairs $(S_i,D_i)$,
thus forming five 
learners.

\begin{figure}
 \centering
    \begin{subfigure}[b]{0.48\columnwidth} 
        \centering
        \includegraphics[width=1.06\columnwidth]{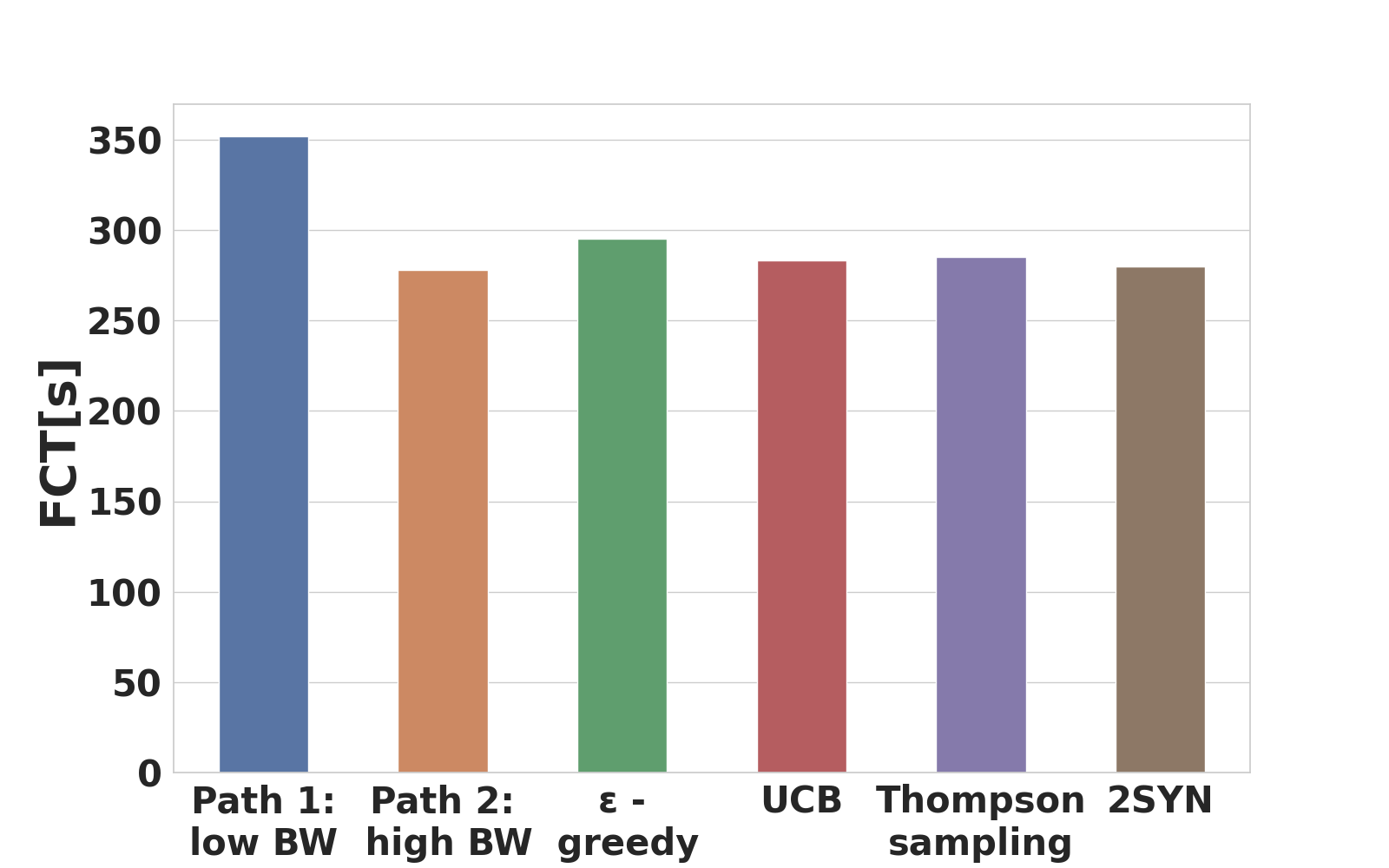}
        \caption{Without bandwidth drop} 
        \label{subfig:simple_mab}
    \end{subfigure}
    \begin{subfigure}[b]{0.48\columnwidth}
        \centering      
        \includegraphics[width=1.06\columnwidth]{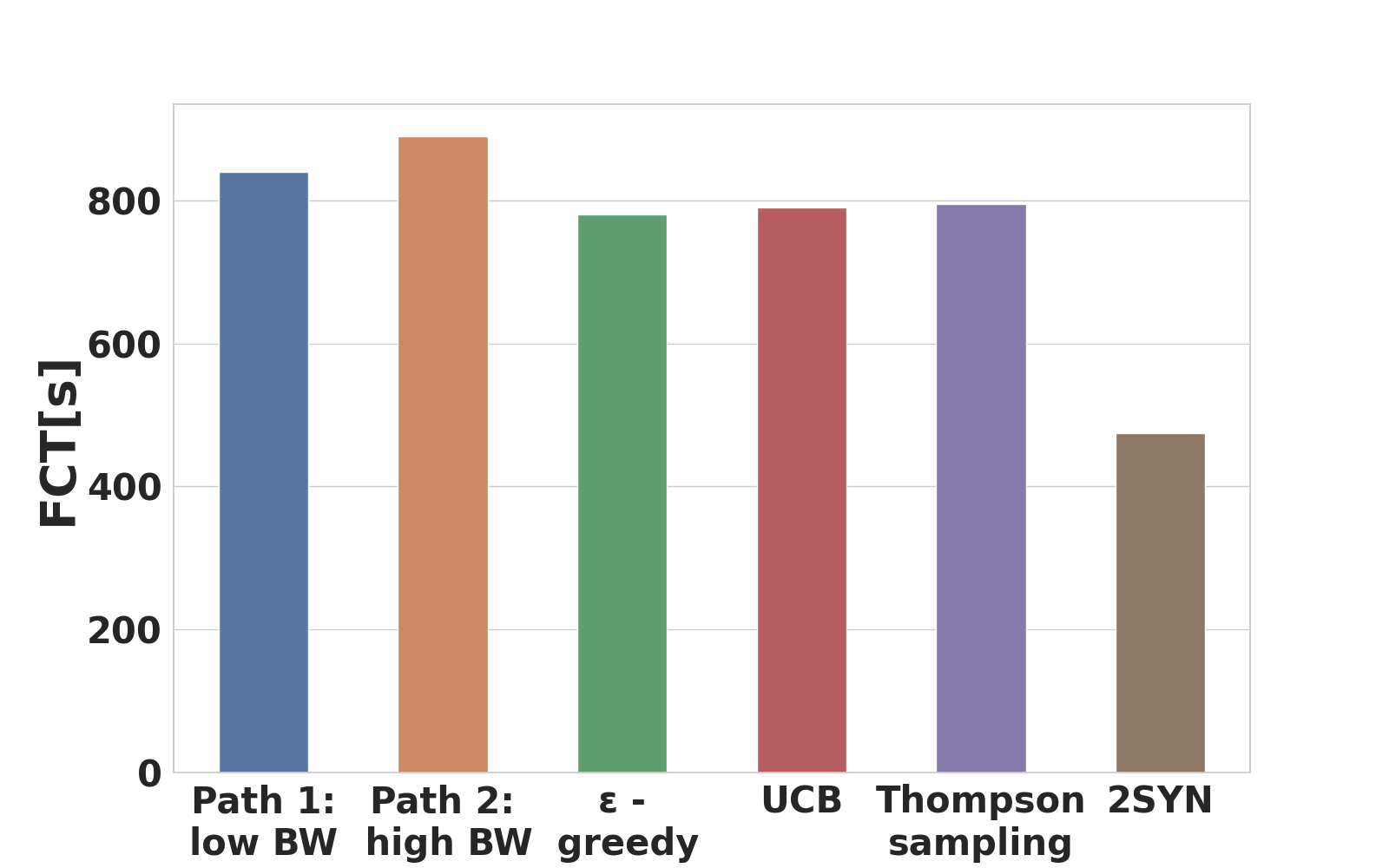}
             \caption{With bandwidth drop}
             \label{subfig:mab_drop}
     \end{subfigure}

     \caption{\textit{\myalg vs. MAB ML algorithms.} 
     (a)~With constant settings, MAB algorithms learn the best path and achieve results similar to \name. 
     (b)~When bandwidth drops, \myalg outperforms MAB algorithms since it adapts faster.
     }
    \label{fig:mab}
    \ReduceVSpace 
\end{figure}

\T{Fixed settings.} At first, for each pair, we send 10-MB files 50 times in a row. After a short time, the MAB algorithms correctly classify the path with the lowest bandwidth.
\fig{subfig:simple_mab} illustrates 
how they converge to near-optimal results. 

\T{Bandwidth drop.} We now set the same initial bandwidth, but after 
$40$\% of the files are sent, we drop the bandwidth of path 2 from $300$\,Mbps to $100$\,Mbps. We send 20 files in a row, with a size of $200$\,MB per file. \fig{subfig:mab_drop} shows that  \myalg clearly outperforms the MAB algorithms. The main reason is that the MAB algorithms base each decision on the path history, and therefore are less resilient to changes in the path conditions, while \myalg adapts much more quickly.
Real-time congestion-aware observations outperform history-based learning.

\section{Conclusion}\label{sec:conclusion}


This paper has introduced \myalg, a lightweight congestion-sensitive algorithm
that can help companies efficiently manage their multi-homed networks. 
It has presented a Linux implementation of \myalg. Then, using lab and real-world experiments, it has shown how \myalg can adapt to various network conditions and outperform alternative algorithms.

\TR{
 \let\thefootnote\relax\footnotetext{
  \Journal{
    \textbf{[Add this 1st paragraph in 1st journal submission only:]} This paper was presented in part at IEEE Infocom '15, Charleston SC, April 2015. Additions to the conference version include new theorems, complete proofs that were previously omitted for space reasons, and additional simulation results.
  }

 G. Student and I. Keslassy are with the Department of Electrical Engineering, Technion, Israel (e-mails: \{grad@tx, isaac@ee\}.technion.ac.il).
 }
}

\ifblind
\else
  \ifacmart
\begin{acks} 
          We would like to thank the many people whom we consulted for this paper,
          including ..., as well as our
          shepherd, ..., and our anonymous reviewers.\\
          This work was partly supported by... 
\end{acks}
   \else
        \section*{Acknowledgment}
        {Many thanks to Aran Bergman and Tal Mizrahi for their useful comments.}
        \new{This work was partly supported by the Israel Science Foundation (grant No. 1119/19), the Hasso Plattner Institute Research School, and the Louis and Miriam Benjamin Chair in Computer-Communication Networks.}
  \fi 
\fi 



\ifacm 
  
  \ifacmart 
      \bibliographystyle{ACM-Reference-Format}
      \bibliography{bib/mybib}
  \else
        \bibliographystyle{abbrv}
        \bibliography{bib/mybib}
  \fi
\else
	\ifusenix 
    	{\footnotesize 
        \bibliographystyle{acm}
        \bibliography{bib/mybib}
        }
      \else 
        \bibliographystyle{IEEEtran}
        \bibliography{IEEEabrv,bib/mybib}
	\fi 
\fi 

\end{document}